**Temperature-independent ferromagnetic resonance shift in Bi doped YIG garnets through magnetic anisotropy tuning.**

**Diane Gouéré[1], Hugo Merbouche[1], Aya El Kanj[1], Felix Kohl[1], Cécile Carrétéro[1], Isabella Boventer[1], Romain Lebrun[1], Paolo Bortolotti[1], Vincent Cros[1], Jamal Ben Youssef[2], Abdelmadjid Anane[1]**

[1] Unité Mixte de Physique CNRS, Thales, Université Paris-Saclay, 91767, Palaiseau, France

[2] Lab-STICC, UMR 6285 CNRS, Université de Bretagne Occidentale, 29238, Brest, France

**Abstract :**

Thin garnet films are becoming central for magnon-spintronics and spin-orbitronics devices as they show versatile magnetic properties together with low magnetic losses. These fields would benefit from materials in which heat does not affect the magnetization dynamics, an effect known as the non-linear thermal frequency shift. In this study, low damping Bi substituted Iron garnet (Bi:YIG) ultra-thin films have been grown using Pulsed Laser Deposition. Through a fine tuning of the growth parameters, the precise control of the perpendicular magnetic anisotropy allows to achieve a full compensation of the dipolar magnetic anisotropy. Strikingly, once the growth conditions are optimized, varying the growth temperature from 405 °C to 475 °C as the only tuning parameter induces the easy-axis to go from out-of-plane to in-plane. For films that are close to the dipolar compensation, Ferromagnetic Resonance measurements yield an effective magnetization $\mu_0 M_{eff}$ (T) that has almost no temperature dependence over a large temperature range (260 K to 400 K) resulting in an anisotropy temperature exponent of 2. These findings put Bi:YIG system among the very few materials in which the temperature dependence of the magnetic anisotropy varies at the same rate than the saturation magnetization. This interesting behavior is ascribed phenomenologically to the sizable orbital moment of $Bi^{3+}$.

diane.gouere@cnrs-thales.fr

madjid.anane@universite-paris-saclay.fr

## I. INTRODUCTION

The recent discoveries in the field of insulating spintronics [1] have triggered an interest for the growth of ultra-thin insulating film based on garnets [2]. The ultra-fast magnetic domain walls (DW) driven by spin-orbit-torques in Bi doped YIG films [3] ($Bi_xY_{3-x}Fe_5O_{12}$) as well as the coherent emission of spin-waves (SW) [4] has made such materials a favorable ground for new physical phenomenon to be explored. Even if garnets have been investigated since the 60's, the emergence of new growth approaches such as Pulsed Laser Deposition (PLD) and off-axis sputtering [5] have proven that garnet ultra-thin films can possess dynamical properties comparable to that of micron-thick films traditionally grown using Liquid Phase Epitaxy (LPE) [6]. Interface transparency to pure spin currents that take place in adjacent heavy metal layers (most often Pt) has allowed for efficient modulation of the SW life-time through the Damping-like spin-orbit-torque [7]. For instance, a fivefold increase of the SW attenuation length in 20 nm thick YIG films has been observed and even auto-oscillations of the magnetization have been reached [8]. Nevertheless, the required electrical current flow for the generation of the torques in Pt also induces a Joule heating in the Pt/YIG bilayers. As a result, detrimental effects on the spin-wave spectrum appear, since heating-induced reduction of the saturation magnetization in turn generates a self-localization of the induced magnetization dynamics that inhibits SW emission and propagation [9]. This non-linear frequency shift is reminiscent, even if the physical origin is different, of the solitons formation in all-metallic spin-Hall nano-oscillators [10]. A solution to this issue has been proposed using the out-of-plane magnetic anisotropy of $Bi_xY_{3-x}Fe_5O_{12}$ system [11]. In parallel to these lines of investigations, the study of spin-orbit-torque induced magnetic motion of DW in garnets thin films (such as $Tm_3Fe_5O_{12}$ or $Bi_xY_{3-x}Fe_5O_{12}$) has pointed out the existence of a weak, but necessary, interfacial Dzyaloshinskii-Moriya Interaction(DMI) [12]. The $Bi_xY_{3-x}Fe_5O_{12}$ system hence appears as being the most versatile material platform as it combines tunable magnetic anisotropy and ultra-low magnetic losses [13]. A precise control of this system will eventually lead to practical magnon-spintronics devices that could perform, for instance, in the static regime for memory application or in the dynamical regime for spin-waves based information processing.

Up to now there are no reports establishing a clear correlation between growth parameters and magnetic anisotropy of doped garnets. Indeed, the magnetic anisotropy parameter for garnets, which is of importance for magnon-spintronics applications, is more usually tuned by extrinsic approaches such as the nature of the substrate, the nature of the dopant atom, or the content of the dopant atom [5,6,13]. In this study, we investigate the temperature dependence of the magnetic anisotropy of low-loss Bi substituted thin YIG films grown by PLD. First, the growth of thin films with optimal structural and magnetic properties has been optimized. Then we show that by tailoring the growth temperature and only the growth temperature, the film magnetic anisotropy can be tuned from out-of-plane to in-plane, yielding a Ferromagnetic Resonance (FMR) frequency's temperature shift that is also tunable. We demonstrate that the FMR response can be finely adjusted and that we can even reach the conditions for which no temperature dependence on a large temperature range (260 K to 400 K) is observed. This provides a route for material properties engineering to solve a long standing issue of SW devices that up to now needed temperature regulation to perform correctly. By using suitable Bi:YIG, such devices can be made much more compact and less bulky by relieving the temperature stability constraints. Furthermore, temperature independent magnetization dynamics will be instrumental for the development of future spin-orbitronics and magnonics radiofrequency processing devices.

## II. EXPERIMENTS

Thin films have been grown with a PLD process using a tripled frequency YAG laser ($\lambda$ = 355 nm) pulsed at 2.5 Hz. We use (111)-oriented substituted-$Gd_3Ga_5O_{12}$ (sGGG) substrate and a single $Bi_1Y_2Fe_5O_{12}$

(Bi:YIG) target. The lattice parameters are a = 12.497 Å for the sGGG substrate and a = 12.45 Å for the Bi:YIG target. The laser fluence (1.3 J.cm$^{-2}$), target-substrate distance (44 mm), oxygen pressure, growth rate, and all the other growth parameters have first been optimized. The oxygen pressure was set at 0.26 mbar during the growth process and Bi:YIG films were grown at temperatures ranging from 405 °C to 475 °C at a growth-rate of 0.41 Å.s$^{-1}$. Note that this temperature range is lower than what most studies report for the growth of garnets. Going from YIG, for which the growth temperature is reported to be superior to 650 °C, to Bi:YIG has indeed an effect on the value of the optimal growth temperature. To preserve 2D like RHEED patterns, the growth temperature needs to be lowered.

The films thicknesses have been determined by X-Ray-Reflectivity (XRR) and the structural characterizations have been performed by X-Ray-Diffraction (XRD) using a 5-axes Empyrean Panalytical set-up with a Kα$_1$ monochromator (λ = 1.54056 Å). The static magnetic properties have been characterized by Magneto-Optical Kerr Effect (MOKE) measurements with a magnetic field applied perpendicular or parallel to the film plane. In addition, we employ the « Polar-mode » to image the domain configuration for our samples with Perpendicular Magnetic Anisotropy (PMA) and the « Pure Transversal-mode » for films with in-plane easy magnetization axis. The characterization of the magnetization dynamics has been done using broadband FMR with frequencies ranging from 4 up to 40 GHz. The FMR measurements have been performed between 260 K and 400 K in order to probe the temperature dependence of the magnetic properties.

All the elaborated films have a thickness of 22 ± 2 nm with a typical root-mean square (RMS) roughness of 0.4 nm over 5x5 µm$^2$ (see [14]). The present study focuses on a series of six films in which a clear correlation is established between the growth temperature and the temperature dependence of the magnetic properties, such as magnetic anisotropy, effective magnetization, and non-linear frequency shift. We use the Kerr-effect to record the hysteresis cycle of the films. In Figure 1, we show that films A to D grown at the lowest temperatures (400 °C < T$_{Growth}$ < 440 °C) exhibit an out-of-plane easy magnetization axis with characteristic loop-shapes and low saturation fields (< 10 mT). Insets of the Polar-Kerr images confirm the uniaxial magnetic anisotropy; they show meander-like domains after demagnetization with characteristic domain size that decreases when increasing the growth temperature. For films E and F grown at higher temperature (T$_G$ = 460 °C and T$_G$ = 475 °C) the hysteresis cycles have been obtained using an in-plane magnetic field in the Pure-Transversal-Kerr mode, as their saturation in the Polar field configuration is reached at much higher magnetic fields than films A-D. These films E-F have an in-plane easy magnetic axis and, as shown by the insets in Figure 1, highly contrasted in-plane domains with zigzag boundaries. These results indicate that modifying the growth temperature alone allows changing the PMA. Easy magnetization axis is found to switch from out-of-plane to in-plane at T$_G$ ≈ 450 °C.

To characterize more precisely the magnetic anisotropy, we use broadband FMR in the field-in-plane configuration. Measurements have been performed from 4 to 20 GHz at room temperature. The Kittel formula is used to extract the value of the effective magnetization defined as $M_{eff} \stackrel{\text{def}}{=} M_s - H_u$ where $M_s$ is the saturation magnetization and $H_u$ is the uniaxial out-of-plane anisotropy field. The frequency dependence ($f_{res}$) of the FMR resonance field ($H_{res}$) is fitted for each film to :

$$f_{res} = \mu_0 \gamma \sqrt{H_{res}(H_{res} + M_{eff})} \qquad \text{(Equation 1)}$$

where $\gamma$ is the electron gyromagnetic ratio, fixed to its free-electron value : 28 GHz/Tesla.

In Figure 2a, we show that the room temperature effective magnetization ($\mu_0 M_{eff}$) for Bi-doped YIG thin films can vary over hundreds of mT (from -155 mT to 220 mT) when the target composition, the substrate, the growth temperature, the laser fluence and the oxygen pressure are varied [13] (films labeled 1, 2 and 3). However, for the samples of interest here, labeled from A to F, we kept all parameters constant except for the growth temperature indicated on the x-axis (Table 1). It is observed for these films that the effective magnetization value changes within a much smaller range (from - 46 mT to +33 mT) and increases almost linearly with $T_G$ (the red line is a guide for the eye). Films grown below $T_G$ = 450 °C have a negative effective magnetization. For films E and F, the effective magnetization is positive. The compensation ($M_{eff} = 0$) for which we observe the out-of-plane to in-plane spin reorientation transition is reached for $T_G \approx 428$ °C ($\mu_0 M_{eff} = -1.5$ mT). These results obtained by FMR are hence consistent with those obtained by Kerr-imaging shown in Figure 1: only films that show meander domains and out-of-plane easy magnetization axis have negative effective magnetizations. The inset in Figure 2a displays the peak-to-peak resonance linewidth values of films A to F. Note that the Gilbert damping parameters are comparable for all films and is about $\alpha = 7.10^{-4}$ at room temperature. Optimization in the measurement geometry could eventually lead to a lower value such that for that of Film F : $\alpha = 4\ 10^{-4}$ (see [14]). The absorption FMR spectra of films E (blue, Figure 2b) and B (red, Figure 2c) are reported for f = 12 GHz: both the peak-to-peak linewidth ($\mu_0 \Delta H_{pp}$) and the resonance field ($\mu_0 H_{res}$) are indicated. We note that the resonance fields are shifted from $\mu_0 H_{res} = 413.1$ mT to $\mu_0 H_{res} = 439.2$ mT when the growth temperature decreases by 40 °C. Moreover, the resonance fields are lower (respectively higher) than the expected resonance field $\mu_0 H_{res}(M_{eff} = 0) = 428.6$ mT calculated using Equation 1 for compensated thin films. The growth temperature thus appears as a control knob for fine tuning of the perpendicular magnetic anisotropy in a continuous way.

In order to obtain a more thorough characterization of the perpendicular magnetic anisotropy properties, FMR measurements from 5 to 40 GHz have been performed in an extended range of temperature (from 260 K to 400 K). These measurements were done for thin films A-F, but also for films reported on Figure 2a (films 1, 2, and 3). These films are presented to illustrate the extent on which the effective magnetization can vary in the Bi:YIG system, they have been grown using a different set of growth parameters (see Table 2). In Figure 3a, we show effective magnetization as a function of the measurement temperature ($T_{MEAS}$) for thin films 1 and 3 with strong out-of-plane anisotropy. At $T_{MEAS}$ = 260 K, $\mu_0 M_{eff}$ values are $-156$ mT and $-102$ mT respectively. As reported in Figure 3b, the effective magnetization gradually increases until $T_{MEAS}$ = 400 K (to $\mu_0 M_{eff} = -127$ mT and $\mu_0 M_{eff} = -77$ mT respectively). The opposite behavior is observed for film 2 (with in-plane anisotropy) for which $\mu_0 M_{eff}$ decreases from $+232$ mT at $T_{MEAS}$ = 260 K to $+166$ mT at $T_{MEAS}$ = 400 K. This range of variation is to be compared to what is measured on films C and F that are closer to compensation, for which $\mu_0 M_{eff}$ varies only by +6 mT and -6 mT between 260 K and 400 K. We note that the absolute thermal shift value is comparable for films C and F (+0.039 mT.K$^{-1}$ and -0.046 mT.K$^{-1}$, respectively). It is observed that films with the largest $|\mu_0 M_{eff}|$ values have large temperature dependence of their effective magnetization. But the temperature independence of $|\mu_0 M_{eff}|$, observed for films C and F, can only be explained by a specific power law between $M_s$ and $K_u$ (see Discussion and [14]). Absorption FMR spectra acquired at 260 K, 300 K and 340 K are shown in Figure 3c for the film closest to compensation (film C) for which the resonance field shift ($\mu_0\ \delta\ H_{res}$) in this range of temperature is only $\mu_0\ \delta\ H_{res} \approx +4$ mT. It is also observed that the absorption linewidth is sharper at $T_{MEAS}$ = 400 K than at lower temperatures as usually seen in thin garnet films [15].

X-Ray Diffraction (XRD) has been used to investigate the structural properties of the Bi:YIG thin films. Inset in Figure 4 displays the entire diffraction spectra (2θ angle varies from 10 to 130 °) of film E in which several orders of diffractions are observed. The ones associated to sGGG substrate are identified for (444) and (888) together with forbidden reflections, (222) and (666). We observe a systematic shoulder on the side of (444) and (888) peaks that is present even on the bare substrate originating, most probably, from a secondary sGGG phase. As shown in Figure 4, the substrate's (888) diffraction peak is at $2\theta_{888}$ = 117.3 ° and its shoulder $2\theta_{888}$ = 117.7 °; Bi:YIG films diffract at $2\theta_{444}$ = 50.8 ° and $2\theta_{888}$ = 118.5 °. Diffraction spectra of thin films grown between $T_G$ = 405 °C and $T_G$ = 475 °C are plotted in the main panel of Figure 4, $2\theta_{888}$ = 118.5 ° corresponds to a pseudo-cubic out-of-plane lattice constant of 1.242 nm. The relaxed $Bi_1Y_2IG$ unit cell parameter is expected to be 1.245 nm (against 1.249 nm for sGGG). As a consequence the 2θ-ω scan points toward tensile strained Bi:YIG films at all growth temperatures. It is shown that the diffraction peaks of all films perfectly overlap leading to a growth-temperature independent strain for Bi:YIG thin films. The identical shape of the films peaks also indicated that the Bi content is constant among all films. The presence of Laue fringes at the (444) reflection is reminiscent of a pseudomorphic growth.

### III. DISCUSSION

One of the important results is that the magnetic anisotropy in Bi:YIG system has two peculiarities, that up to our knowledge have not been previously reported. First it can be finely tuned by the variation of the growth temperature. Second, its temperature dependence has a non-usual behavior in the vicinity of the dipolar compensation ($\mu_0 M_{eff} \approx 0$). In Bi doped YIG the out-of-plane magnetic anisotropy has two origins: a magneto-elastic and a growth induced term. The former results from epitaxial strain, for which the pseudomorphic in-plane elastic deformation yields: either a negative or a positive magneto-strictive effect. For garnets deposited along the (111) axis, the magneto-strictive parameter ($\lambda_{111}$) is negative. A compressive strain favors in-plane magnetic anisotropy, while a tensile strain favors an out-of-plane easy magnetization axis. In this study with a doping level of x = 1 ($Bi_xY_{3-x}Fe_5O_{12}$) and the sGGG substrate, we are in the former case i.e. a tensile strain as the relaxed parameter ($a_0$) of Bi:YIG is smaller than that of the substrate. Hence using the magneto-elastic theory, the out-of-plane misfit $\Delta a = a_{sub} - a_0$ (where $a_{sub}$ is the lattice parameter of the substrate) is positive and yields a positive magneto-elastic anisotropy constant ($K_{MO}$) [16,17]:

$$K_{MO} = -\left[\frac{3}{2} \times \frac{E}{1+\mu} \times \left(\frac{\Delta a}{a_0}\right) \times \lambda_{111}\right] \text{ (Equation 2)}$$

$$\text{and } a_0 = \left[\frac{(a^\perp_{film} - a_{sub})}{1+\mu} \times (1-\mu) + a_{sub}\right] \text{ (Equation 3)}$$

Where $a^\perp_{film}$ = 1.242 nm is the experimentally obtained out-of-plane lattice parameter, $E = 2.055 \times 10^{11}$ J.m$^{-3}$ and $\mu$ = 0.29 are the Young modulus and the Poisson coefficient respectively, and $\lambda_{111} = -5.0575 \times 10^{-6}$ is the magneto-strictive parameter [18,19]. Considering the XRD measurements performed on the Bi:YIG thin films, we calculated $a_0$ = 1.245 nm, $\Delta_a = 4.7 \times 10^{-3}$ nm and $K_{MO} = 4.56 \times 10^3$ J.m$^{-3}$ which favors an out-of-plane anisotropy. This value of $K_{MO}$ is constant for all films of the study (A, B, C, D, E, and F) as no change in the (888) diffraction peak is observed (see Figure 4). For Bi:YIG thin films, the film's saturation magnetization ($M_S$) has been measured and its value is $M_S$ = 116 kA.m$^{-1}$ ($\mu_0 M_S$ = 145.8 mT) as previously reported by Lin *et al.* [20]. Note that this value is lower than that of un-doped YIG thin films for which $M_S$ = 140 kA.m$^{-1}$. We can then deduce the magneto-elastic anisotropy field: $\mu_0 H_{MO} = \mu_0 \frac{2 K_{MO}}{M_S}$ = 98 mT. This value is roughly 2/3 of $\mu_0 M_S$, alone it is yet far from sufficient to overcome the shape anisotropy. One needs to consider an extra-term introduced five decades ago by H. Callen [21], known as the growth induced anisotropy. It arises from the preferential occupation by Bi atoms of nonequivalent dodecahedral sites of the unit cell resulting in a symmetry lowering along the growth direction. This term contributes to the magnetic free energy of the system as $K_{Growth} \sin^2 \theta$, where θ is the polar angle relative to surface normal and $K_{Growth}$ the associated anisotropy constant. $K_{Growth}$ is positive under either a compressive or a tensile strain. In the present case, the fact that a vanishing effective magnetization is found, implies that $K_{Growth}$ is of the same order than $K_{MO}$. Hence, the sum of both contributions compensates the dipolar term, whereas the cubic magnetic anisotropy is more than an order of magnitude smaller [16] and can be neglected at first order.

Our first main observation is the correlation between T$_{Growth}$ and the magnetic anisotropy. The relevant question is thus to identify which one of the anisotropy terms is varying with the growth temperature (T$_{Growth}$). From X-ray diffraction, we obtain that the (888) diffraction peak remains identical for all films. Assuming a pseudomorphic growth for films A to F, we conclude that the relaxed cell parameter (a$_0$) is identical for all films. This has two consequences: (*i*) the Bi content and (*ii*) the strain state are identical for films A to F. We can therefore discard any significant variation of the magneto-elastic anisotropy among these films. As a consequence the only remaining anisotropy term affected by T$_{Growth}$ is the growth induced anisotropy. In fact, we deduce a very strong variation of K$_{Growth}$ than span from 4327 J.m$^{-3}$ to 667 J.m$^{-3}$ over a growth temperature range of only 75 °C. Such a large effect of T$_{Growth}$ on K$_{growth}$ is largely unexpected. We speculate that the effect of growth temperature is due to a change in the Bi atoms distribution over the nonequivalent six dodecahedral sites [22]. Yet difficult to evidence experimentally, this explanation is supported by the fact that increasing T$_{Growth}$ induces a monotonic decrease of growth anisotropy as expected from entropy arguments.

The second main observation is that the effective magnetization can be tuned to be almost temperature independent over a wide temperature range for films close to dipolar compensation, yielding an FMR frequency that is stable within 100 MHz ( $\Delta f \approx \frac{\gamma \times \Delta M_{eff}}{2}$ ) over a 140 K temperature range. As a comparison, pure YIG films grown by PLD with similar thickness, exhibits about an order of magnitude larger frequency shift over the same temperature range. Such a temperature dependence of the effective magnetization is the second peculiarity of the films under investigation. To have this effect, having a small value of $|\mu_0 M_{eff}|$, yet necessary, is not the only condition. In the system under investigation, $H_u$ and $M_s$ should have the same temperature dependence to keep $|\mu_0 M_{eff}|$ constant. Callen and Callen have proposed the following power law to relate the temperature dependence of the anisotropy to the temperature dependence of the magnetization [23] :

$$\frac{K_u(T)}{K_u(0)} = \left[\frac{M_s(T)}{M_s(0)}\right]^n \text{(Equation 4)}$$

with $K_u$ is the uniaxial magnetic anisotropy, $M_s$ the saturation magnetization and $n$ an exponent that is related to the crystal symmetry. It has been predicted by Van Vleck and Zener to be $n = 3$ in case of uniaxial anisotropy [24,25]. In our case, for a film with vanishing effective magnetization, we observe that $M_{eff}(T) \approx 0$, i.e. $M_s(T) \approx H_u(T)$. The uniaxial anisotropy field is therefore equal to the saturation magnetization at first order, for all explored temperatures (from 260 K to 400 K). Using Equation 4, the uniaxial magnetic anisotropy $K_u(T) = \frac{\mu_0}{2} M_s(T) \times H_u(T)$ is thus simply proportional to the square of the saturation magnetization. This corresponds to an exponent $n = 2$ over a temperature range spanning at least from 260 K to 400 K, while the films Curie temperature is expected to be 559 K. A similar exponent $n = 2$ has been found while plotting $K_u$(T) as a function of $M_s$(T) and fitting the slope with Equation 4, (Fig. 3d) (see [14]). The authors want to stress out here that such exponent is not expected by the theory that has been developed for many system including garnets. As a matter of fact, many systems in the literature deviate from their theoretical n exponent (given for cubic or uniaxial anisotropy). Indeed interestingly, there are few systems that deviate from the standard value of $n = 3$ [26–28]. Chatterjee *et al.* found an exponent $n = 2.6$ for the NiFe$_2$O$_4$ system, Wang *et al.* found an exponent of $n = 1$ for BaFe$_{12}$O$_{19}$, after fitting experimental data. Thus, although the deviation of the existent model is then not specific to Bi:YIG system, the explanation may still differ from one system to the other. However, a value of $n = 2$ has only been reported for ordered Pt-Fe alloys of L1$_0$-type where this specific temperature dependence has been ascribed to a two-ion anisotropy related to the induced moments on the Pt [29]. For our films, even if $Bi^{+3}$ ions in Bi:YIG are not expected to carry a spin magnetic moment, M1-edge X-ray Circular magnetic dichroism (XMCD) experiments have unambiguously revealed that Bi atoms bear a sizable orbital moment [30]. This explains the fact that we observe a significantly smaller magnetization saturation than that of YIG confirming previous observations (as Vertruyen *et al.* [31]). Additionally, the presence of the orbital moment may also explained a two-ions like magnetic anisotropy behavior in Bi:YIG that would then be related to the complex [5d,6p] Bi orbitals hybridization with Fe. We hence conclude that our observations are resulting from two effects that are growth induced anisotropy sensitivity to growth temperature and the magnetic state of Bismuth. A first-principal investigation could give enlightening insights on the inter-links between these two effects, which is beyond this work.

## IV. CONCLUSION

In conclusion, we show *(i)* how to finely tune the magnetic anisotropy in garnets and *(ii)* how to obtain a temperature-free variation of their uniform magnetization dynamics characteristics when achieving a vanishing effective magnetization while preserving a low damping ($\alpha \approx 7 \cdot 10^{-4}$). We explain these observations, based on the growth-induced anisotropy for the former and on the importance to consider Bismuth orbital-momentum for the latter. Growth induced anisotropy is found to be very sensitive to growth temperature whereas no significant effect is observed on the magneto-elastic anisotropy. The anisotropy temperature exponent is found to change with the anisotropy; its value is $n \sim 2$ when the out-of-plane anisotropy compensates the shape anisotropy. The practical implications of these findings for the field of magnonics and spin-orbitronics are numerous. At first, they solve a long standing issue of YIG based radiofrequency devices [32], that most often need thermally regulated packaging. Given the much smaller thermal sensitivity of optimized Bi:YIG, such technical constraints can now be lifted and more compact and less power hungry devices could be developed. Furthermore, we explain why using Bi:YIG has been very important for the physics of spin-orbit-torque in magnonics, where Joule heating is unavoidable. Not only compensating exactly the dipolar contribution is necessary to avoid non-linear effects such as non-linear magnon-magnon interactions [4], but it is equally important to suppress any temperature dependence of FMR frequency to avoid spin wave localization. As the only temperature dependent term in Equation 1 is the effective magnetization, for low damping materials, such condition can be reached in Bi:YIG and as far as the authors are aware of, only in the Bi:YIG system. Finally, we provide perspectives on how magnons dispersion relation can be engineered to have the desired temperature dependence. Such latitude can open a new field for magnonics lenses where the magnetic medium has reconfigurable effective indexes for spin waves optical elements [33].

**ACKNOWLEDGMENTS**

We acknowledge support from the French National Research Agency (ANR), ANR MARIN ANR-20-CE24-0012. This project has received funding from the European Union's Horizon 2020 research and innovation program under the Marie Skłodowska-Curie grant agreement No 861300. and under FET-Open Grant No. 899646 (k-NET), and as part of the "Investissements d'Avenir" program (Labex NanoSaclay, reference: ANR-10-LABX-0035 "SPiCY").

**References**

[1] Y. Kajiwara, K. Harii, S. Takahashi, J. Ohe, K. Uchida, M. Mizuguchi, H. Umezawa, H. Kawai, K. Ando, K. Takanashi, S. Maekawa, and E. Saitoh, *Transmission of Electrical Signals by Spin-Wave Interconversion in a Magnetic Insulator*, Nature **464**, 262 (2010).

[2] G. Schmidt, C. Hauser, P. Trempler, M. Paleschke, and E. T. Papaioannou, *Ultra Thin Films of Yttrium Iron Garnet with Very Low Damping: A Review*, Phys. Status Solidi Basic Res. **257**, (2020).

[3] L. Caretta, S. H. Oh, T. Fakhrul, D. K. Lee, B. H. Lee, S. K. Kim, C. A. Ross, K. J. Lee, and G. S. D. Beach, *Relativistic Kinematics of a Magnetic Soliton*, Science (80-. ). **370**, 1438 (2020).

[4] V. E. Demidov, S. Urazhdin, A. Anane, V. Cros, and S. O. Demokritov, *Spin-Orbit-Torque Magnonics*, J. Appl. Phys. **127**, (2020).

[5] J. Ding, C. Liu, Y. Zhang, U. Erugu, Z. Quan, R. Yu, E. McCollum, S. Mo, S. Yang, H. Ding, X. Xu, J. Tang, X. Yang, and M. Wu, *Nanometer-Thick Yttrium Iron Garnet Films with Perpendicular Anisotropy and Low Damping*, Phys. Rev. Appl. **14**, 1 (2020).

[6] Y. H. Rao, H. W. Zhang, Q. H. Yang, D. N. Zhang, L. C. Jin, B. Ma, and Y. J. Wu, *Liquid Phase Epitaxy Magnetic Garnet Films and Their Applications*, Chinese Phys. B **27**, (2018).

[7] A. Hamadeh, O. D'Allivy Kelly, C. Hahn, H. Meley, R. Bernard, A. H. Molpeceres, V. V. Naletov, M. Viret, A. Anane, V. Cros, S. O. Demokritov, J. L. Prieto, M. Muñoz, G. De Loubens, and O. Klein, *Full Control of the Spin-Wave Damping in a Magnetic Insulator Using Spin-Orbit Torque*, Phys. Rev. Lett. **113**, 1 (2014).

[8] M. Evelt, V. E. Demidov, V. Bessonov, S. O. Demokritov, J. L. Prieto, M. Muñoz, J. Ben Youssef, V. V. Naletov, G. De Loubens, O. Klein, M. Collet, K. Garcia-Hernandez, P. Bortolotti, V. Cros, and A. Anane, *High-Efficiency Control of Spin-Wave Propagation in Ultra-Thin Yttrium Iron Garnet by the Spin-Orbit Torque*, Appl. Phys. Lett. **108**, 20 (2016).

[9] V. E. Demidov, S. Urazhdin, G. de Loubens, O. Klein, V. Cros, A. Anane, and S. O. Demokritov, *Magnetization Oscillations and Waves Driven by Pure Spin Currents*, Phys. Rep. **673**, 1 (2017).

[10] T. Chen, R. K. Dumas, A. Eklund, P. K. Muduli, A. Houshang, A. A. Awad, P. Durrenfeld, B. G. Malm, A. Rusu, and Akerman, *Spin-Torque and Spin-Hall*, Proc. IEEE **104**, (2016).

[11] M. Evelt, L. Soumah, A. B. Rinkevich, S. O. Demokritov, A. Anane, V. Cros, J. Ben Youssef, G. De Loubens, O. Klein, P. Bortolotti, and V. E. Demidov, *Emission of Coherent Propagating Magnons by Insulator-Based Spin-Orbit-Torque Oscillators*, Phys. Rev. Appl. **10**, 1 (2018).

[12] S. Ding, A. Ross, R. Lebrun, S. Becker, K. Lee, I. Boventer, S. Das, Y. Kurokawa, S. Gupta, J. Yang, G. Jakob, and M. Kläui, *Interfacial Dzyaloshinskii-Moriya Interaction and Chiral Magnetic Textures in a Ferrimagnetic Insulator*, Phys. Rev. B **100**, 100406 (2019).

[13] L. Soumah, N. Beaulieu, L. Qassym, C. Carrétéro, E. Jacquet, R. Lebourgeois, J. Ben Youssef, P. Bortolotti, V. Cros, and A. Anane, *Ultra-Low Damping Insulating Magnetic Thin Films Get Perpendicular*, Nat. Commun. **9**, 1 (2018).

[14] See Supplemental Material at [Url] for more details on AFM, FMR and explanations about the magnetic anisotropy exponent

[15] N. Beaulieu, N. Kervarec, N. Thiery, O. Klein, V. Naletov, H. Hurdequint, G. De Loubens, J. Ben Youssef, and N. Vukadinovic, *Temperature Dependence of Magnetic Properties of a Ultrathin*

*Yttrium-Iron Garnet Film Grown by Liquid Phase Epitaxy: Effect of a Pt Overlayer*, IEEE Magn. Lett. **9**, (2018).

[16] M. S. Lucile, *Pulsed Laser Deposition of Substituted Thin Garnet Films for Magnonic*, (2019).

[17] P. Hansen and J. P. Krumme, *Magnetic and Magneto-Optical Properties of Garnet Films*, Thin Solid Films **114**, 69 (1984).

[18] J. Ben Youssef, *Elaboration Par Epitaxie En Phase Liquide, Caracterisation et Etude Physique Des Filmsminces de Grenats Ferrimagnetiques Susbstitues Par Des Ions Bismuth*, (1989).

[19] P. Hansen, C.-P. Klages, J. Schuldt, and K. Witter, *Magnetic and Magneto-Optical Properties of Bismuth-Substituted Lutetium Iron Garnet Films*, **31**, (1985).

[20] Y. Lin, L. Jin, H. Zhang, Z. Zhong, Q. Yang, Y. Rao, and M. Li, *Bi-YIG Ferrimagnetic Insulator Nanometer Films with Large Perpendicular Magnetic Anisotropy and Narrow Ferromagnetic Resonance Linewidth*, J. Magn. Magn. Mater. **496**, 165886 (2020).

[21] H. Callen, *Growth-Induced Anisotropy by Preferential Site Ordering in Garnet Crystals*, Appl. Phys. Lett. **18**, 311 (1971).

[22] P. Novak, *Contribution of $Fe^{3+}$ Ions to the Growth Induced Anisotropy in Garnet Films*, Czech. J. Phys. B **34**, (1984).

[23] H. B. Callen and E. Callen, *The Present Status of the Temperature Dependence of Magnetocrystalline Anisotropy, and the l(L+1) 2 Power Law*, J. Phys. Chem. Solids **27**, 1271 (1966).

[24] J. H. Van Vleck, *On the Anisotropy of Cubic Ferromagnetic Crystals*, Phys. Rev. **52**, 1178 (1937).

[25] C. Zener, *Classical Theory of the Temperature Dependence of Magnetic Anisotropy Energy*, Phys. Rev. **96**, 1 (1954).

[26] B. K. Chatterjee, C. K. Ghosh, and K. K. Chattopadhyay, *Temperature Dependence of Magnetization and Anisotropy in Uniaxial $NiFe_2O_4$ Nanomagnets: Deviation from the Callen-Callen Power Law*, J. Appl. Phys. **116**, (2014).

[27] J. Wang, F. Zhao, W. Wu, and G. M. Zhao, *Unusual Temperature Dependence of the Magnetic Anisotropy Constant in Barium Ferrite $BaFe_{12}O_{19}$*, J. Appl. Phys. **110**, 10 (2011).

[28] G. Long, H. Zhang, D. Li, R. Sabirianov, Z. Zhang, and H. Zeng, *Magnetic Anisotropy and Coercivity of $Fe_3Se_4$ Nanostructures*, Appl. Phys. Lett. **99**, 4 (2011).

[29] K. Inoue, H. Shima, A. Fujita, K. Ishida, K. Oikawa, and K. Fukamichi, *Temperature Dependence of Magnetocrystalline Anisotropy Constants in the Single Variant State of L 10-Type FePt Bulk Single Crystal*, Appl. Phys. Lett. **88**, 86 (2006).

[30] A. Rogalev, J. Goulon, F. Wilhelm, C. Brouder, A. Yaresko, J. Ben Youssef, and M. V. Indenbom, *Element Selective X-Ray Magnetic Circular and Linear Dichroisms in Ferrimagnetic Yttrium Iron Garnet Films*, J. Magn. Magn. Mater. **321**, 3945 (2009).

[31] B. Vertruyen, R. Cloots, J. S. Abell, T. J. Jackson, R. C. Da Silva, E. Popova, and N. Keller, *Curie Temperature, Exchange Integrals, and Magneto-Optical Properties in off-Stoichiometric Bismuth Iron Garnet Epitaxial Films*, Phys. Rev. B - Condens. Matter Mater. Phys. **78**, 1 (2008).

[32] W. S. Ishak, *Magnetostatic Wave Technology: A Review*, Proc. IEEE **76**, 171 (1988).

[33] M. Vogel, P. Pirro, B. Hillebrands, and G. Von Freymann, *Optical Elements for Anisotropic Spin-Wave Propagation*, Appl. Phys. Lett. **116**, (2020).

**Table 1:** Calculated magneto-elastic anisotropy constant ($K_{MO}$) for films grown between 405 °C and 475 °C, all other growth condition being identical. The deduced relaxed cell parameter($a_0$) is given.

| Film | Laser Fluence (J.cm$^{-2}$) | Growth Temperature (°C) | Growth Rate (Å.s$^{-1}$) | Thickness (nm) | $a_0$ (nm) | $K_{MO}$ (J.m$^{-3}$) |
|---|---|---|---|---|---|---|
| A | | 405 | | | | |
| B | | 420 | | | | |
| C | 1.3 | 428 | 0.41 | 22 | 1.245 | 4560 |
| D | | 442 | | | | |
| E | | 460 | | | | |
| F | | 475 | | | | |

**Table 2:** Growth parameters of thin films 1, 2 and 3 compared with the one of films A and F of the study. The main difference is the laser fluence varying from 1 to 1.7 J.cm$^{-2}$.

| Film | Chemical composition | Growth temperature (°C) | O$_2$ Pressure (mbar) | Thickness (nm) | Laser Fluence (J.cm$^{-2}$) |
|---|---|---|---|---|---|
| 1 | Bi$_1$-YIG//sGGG | 470 |  | 21 | 1.6 |
| 2 | Bi$_{0.7}$-YIG//GGG | 510 |  | 23 | 1.7 |
| 3 | Bi$_1$-YIG//sGGG | 510 | 0.25 | 25 | 1 |
| A | Bi$_1$-YIG//sGGG | 405 |  | 22 | 1.3 |
| F | Bi$_1$-YIG//sGGG | 475 |  | 22 | 1.3 |

**Figures**:

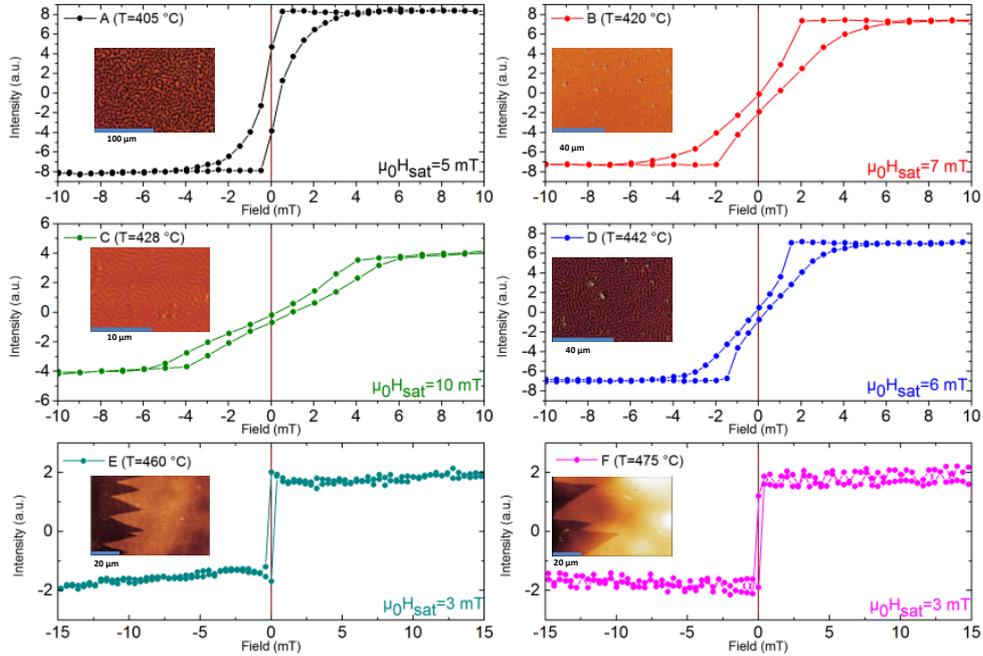

**Figure 1:** Magneto-optical Kerr Effect (MOKE) measurements in Polar-mode (A, B, C, and D) and in the Pure-Transversal mode (E and F) for the 22 nm thick Bi:YIG//sGGG films. Films grown between 405 °C and 442 °C have meander-like domain structure at remanance and an out-of-plane easy magnetization axis. Films E and F grown at 460 °C and 475 °C are in-plane magnetized as evidenced by in-plane hystersesis cycles and the in-plane domains having characterestic zigzag boundaries. The saturation field is also indicated.

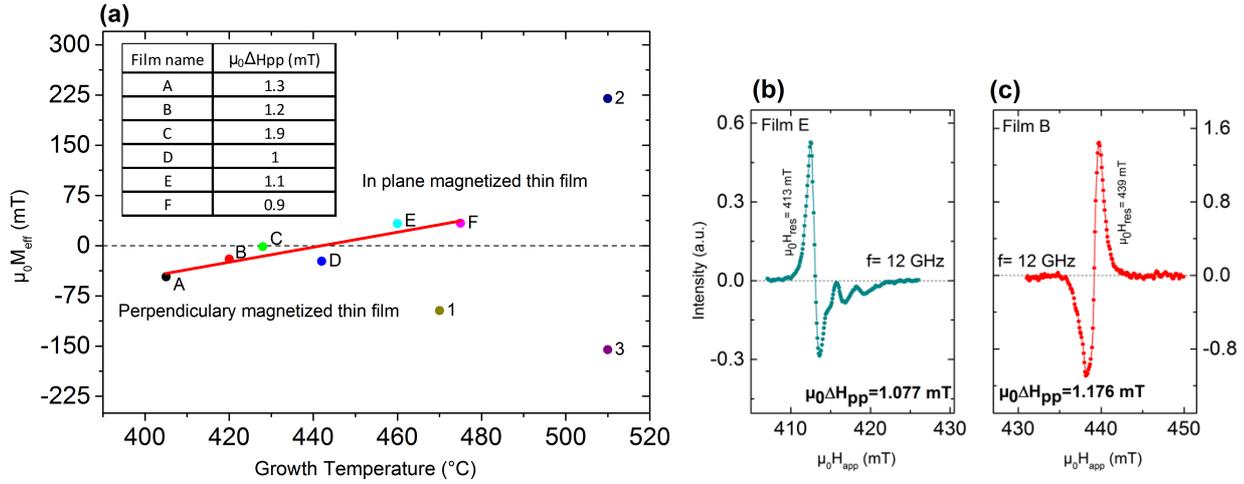

**Figure 2:** Room-temperature Ferromagnetic Resonance (FMR) measurements in the in-plane field configuration. (a) The effective magnetization obtained through fitting to the Kittel formula (Equation 1) is ploted as a function of the growth temperature. The red line is a guide-for-the-eye for the present series of six films (A, B, C, D, E, and F). Note the quasi-linear temperature dependence of the effective magnetization, that is negative for films with out-of-plane easy magnetization axis (A, B, C, and D) and positive for films with in-plane easy magnetization axis (E,F). The linewidth at 12 GHz corresponding to the broadband FMR measurements is reported as an inset. Films labeled 1, 2, and 3 corespond to Bi:YIG films with larger effective magnetizations obtained either by changing the Bi content (1 and 2) or by changing the laser power (3). (b) and (c) FMR absorption spectra for films E and B showing the lower resonance field for in-plane magnetized films.

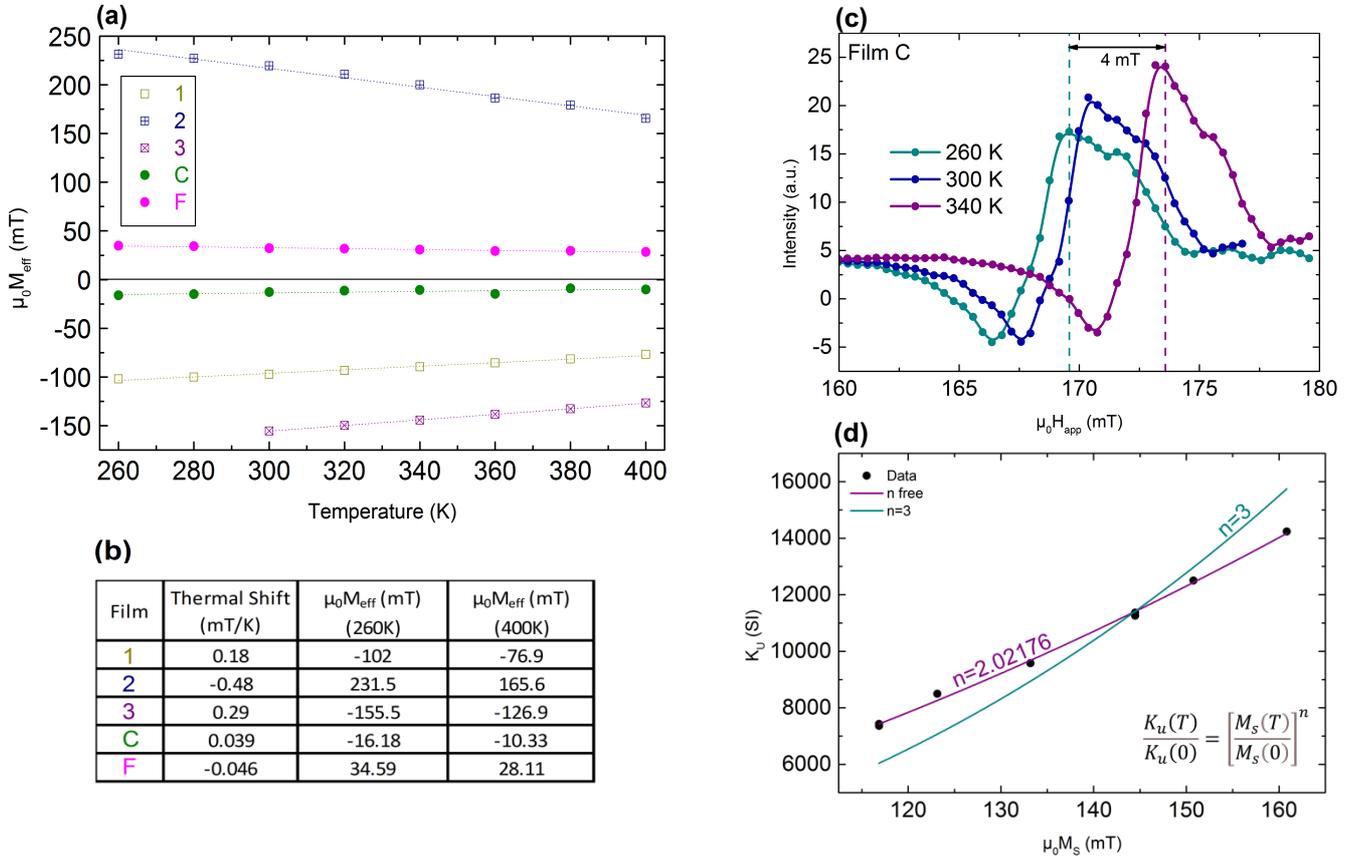

**Figure 3**: Temperature dependence of the Ferromagnetic Resonance (FMR) measurements. (a) The effective magnetization is ploted as a function of the measurement temperature for films (C,F) of the study and three other films (1-3). For films with large effective magnetizations (1, 2, and 3) we observe a significant variation of its value when sweeping the measurement temperature. (b) Whereas for films with much smaller effective magnetization (C,F) the temperature dependence is much smaller. (c) FMR absorption spectra for film C at three measurements temperatures at 5 GHz. The shift in the resonance field does not exceed 4 mT. (d) Inferred dependence of the out-of-plane anisotropy constant as a function of the saturation magnetization, including a fit which corresponds to an anisotropy exponent $n = 2$.

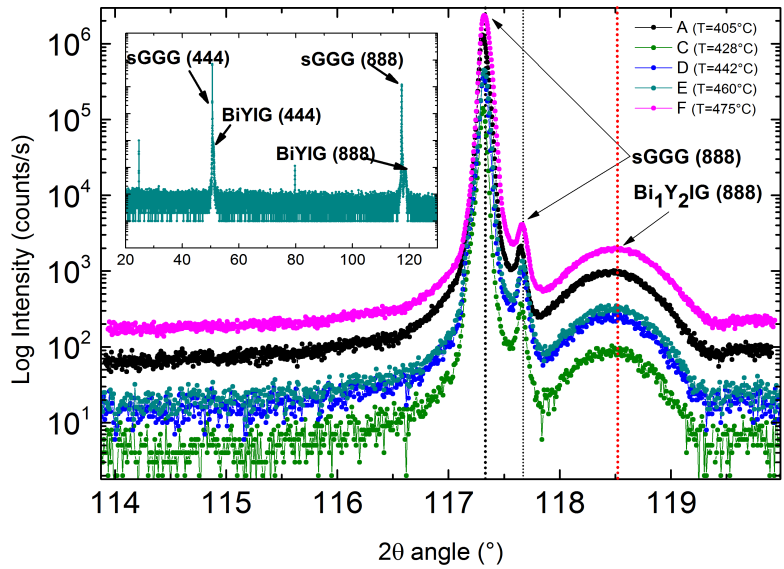

**Figure 4:** θ–2θ X-Ray diffraction (XRD) spectra for Bi:YIG//sGGG. The main panel (with several films grown between 405 °C and 475 °C) in which (888) reflections show that the out-of-plane interlayer distance is identical for all films (curves have been offseted for clarity). The inset displays a typical XRD pattern in 10°-130° range. No parasitic phase is observed.

# Supplementary material

**Temperature-independent ferromagnetic resonance shift in Bi doped YIG garnets through magnetic anisotropy tuning.**

**Diane Gouéré[1], Hugo Merbouche[1], Aya El Kanj[1], Felix Kohl[1], Cécile Carrétéro[1], Isabella Boventer[1], Romain Lebrun[1], Paolo Bortolotti[1], Vincent Cros[1], Jamal Ben Youssef[2], Abdelmadjid Anane[1]**

[1] Unité Mixte de Physique CNRS, Thales, Université Paris-Saclay, 91767, Palaiseau, France

[2] Lab-STICC, UMR 6285 CNRS, Université de Bretagne Occidentale, 29238, Brest, France

## 1. Atomic force microscopy characterization (AFM):

The typical root-mean-square (RMS) roughness of the films is 0.4 nm over a surface of 25 µm². The atomic force microscopy images of sample A, B and F are shown in Figure A. The largest RMS roughness, for sample B, is no more than 0.623 nm. This is more than 30 times lower than the film thickness, we can therefore neglect any shape induced variation of the effective magnetization.

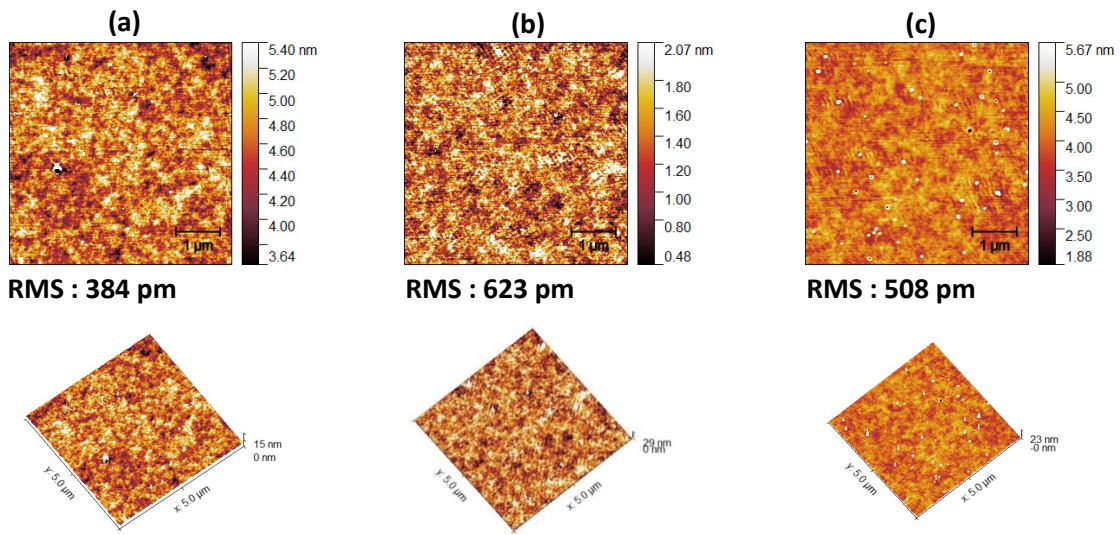

*Figure A: Atomic Force Microscopy (AFM) imaging of (a) film A (b) film B and (c) film F and their 3D view over a surface of 25 µm² recorded in tapping mode. The associated RMS are indicated on the figures.*

## 2. Ferromagnetic resonance characterization (FMR):

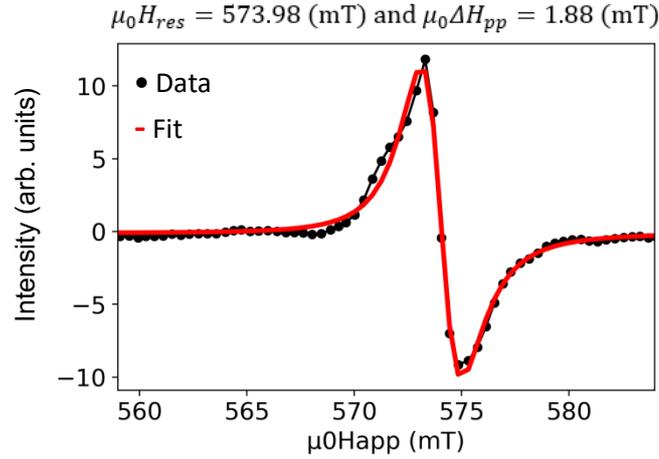

*Figure B: Plot of the derivative of absorption spectra at $f_{res}$=16 GHz for film A. The red line is the fit used to extract both: $\mu_0 H_{res}$ (mT) and $\mu_0 \Delta H_{pp}$ (mT) parameters for each frequency. The film is probed by FMR in the in-plane configuration at room temperature.*

The resonance field ($\mu_0 H_{res}$) and the peak-to-peak linewidth ($\mu_0 \Delta H_{pp}$) are extracted for each resonance frequency ($f_{res}$) from the fits. The Gilbert damping parameter α is fitted using:

$$\mu_0 \Delta H_{pp} = \mu_0 \Delta H_0 + \frac{4\pi \alpha f_{res}}{\gamma \mu_0 \sqrt{3}}$$

The first term corresponds to an inhomogeneous broadening that may be important while measuring PMA thin films (like film A) in the in-plane configuration. The damping parameter α is extracted from the slope when plotting $\mu_0 \Delta H_{pp}(f_{res})$. For films A and F, we show the variation over 8 or 12 GHz:

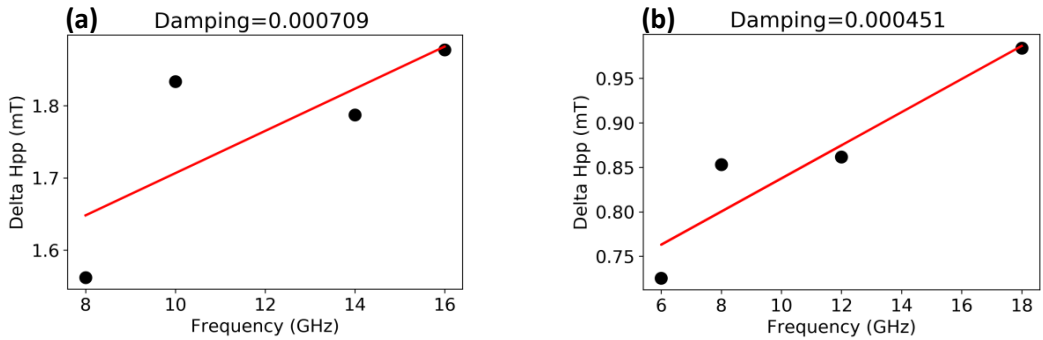

*Figure C: Plot of $\mu_0 \Delta H_{pp}$ versus $f_{res}$ and extracted Gilbert damping parameter for (a) film A where $\alpha = 0.0007$ and (b) film F where $\alpha = 0.00045$. Both films are probed by FMR in the in-plane configuration at room temperature.*

In the main text, we indicated a Gilbert damping of $\alpha = 7 * 10^{-4}$ that is the extracted value for film A with PMA. This value is an upper bound. Optimization in the measurement geometry could eventually lead to a lower value such that for that of film F : $\alpha = 4 * 10^{-4}$.

## 3. Magnetic Anisotropy exponent analysis

Ferromagnetic Resonance (FMR) measurement performed between 260 K and 400 K yields the temperature dependence of the effective magnetization $M_{eff}(T)$ (see Figure 3 in the main), the effective magnetization value of films C and F is stable (temperature coefficient ~ $4.10^{-2}$ mT.K$^{-1}$). We also measure independently the temperature dependence of the magnetization $M_s(T)$ using SQUID magnetometry over the same temperature range. The combination of these two measurements allows to obtain the uniaxial anisotropy constant $K_u(T)$ using: $M_{eff}(T) = M_s(T) - H_u(T)$ and $H_u(T) = \frac{2K_u(T)}{\mu_0 M_s(T)}$

We then plot $K_u(T)$ versus $M_s(T)$ and fit to Equation 1. The result is shown in Figure D (Figure 3d in the main). The best fit for the exponent n is 2.02 very close to the claimed value of 2. This procedure however can be subject to some criticisms. There are additive errors and the dynamic range may be viewed as insufficient even if it allows to easily rule out the n=3 as a possible value (see green curve in Figure D).

$$\frac{K_u(T)}{K_u(0)} = \left[\frac{M_s(T)}{M_s(0)}\right]^n \text{ (Equation 1)}$$

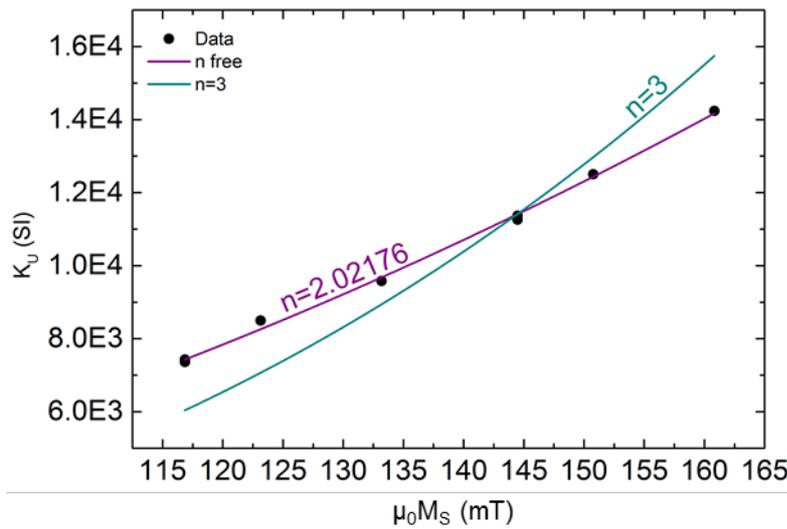

*Figure D: Plot of $K_u(T)$ extracted from the FMR temperature dependent measurement as a function of $M_s(T)$ extracted from SQUID magnetometry measurements. Fits give a n exponent value of n=2.02 close to n=2.*

The second approach relies on the fact that we do observe a vanishing temperature dependence of the effective magnetization :

<u>Temperature scaling argument</u> : We observe that for films that are close to anisotropy compensation:

$$M_{eff}(T) = M_s(T) - H_u(T) \approx 0 \text{ over all the temperature range}$$

As a result:

$$H_u(T) \approx M_s(T) \text{ (Equation 2)}$$

$$\frac{H_u(T)}{H_u(0)} \approx \frac{M_s(T)}{M_s(0)}$$

$H_u(0)$ and $M_s(0)$ refer to the values of the anisotropy field and the saturation magnetization at the lower bound of the temperature range of interest.

By replacing the anisotropy field $H_u(T)$ with the anisotropy constant $K_u(T)$ defined as:

$$H_u(T) = \frac{2K_u(T)}{\mu_0 M_s(T)}$$

We obtain:

$$\frac{K_u(T) \cdot M_s(0)}{M_s(T) \cdot K_u(0)} \approx \frac{M_s(T)}{M_s(0)}$$

And finally:

$$\frac{K_u(T)}{K_u(0)} \approx \left[\frac{M_s(T)}{M_s(0)}\right]^2$$

Identifying this result to Equation 1 yields an exponent of n=2.

Mathematical derivation of the temperature stability condition:

It is worth noting that having a small $M_{eff}$ alone, as it can be obtained by anisotropy engineering is not enough to guaranty a vanishing temperature dependence of $M_{eff}$. This can be mathematically demonstrated within the Callen and Callen model. A vanishing temperature dependence of the effective magnetization can only be obtained for $n = 2$ :

According to the definition: $M_{eff}(T) = M_s(T) - H_u(T)$ (Equation 3)

The uniaxial anisotropy field is linked to $K_u(T)$ and $M_s(T)$ by: $H_u(T) = \frac{2K_u(T)}{\mu_0 M_s(T)}$ (Equation 4)

From Callen and Callen power law we have: $K_u(T) = K_u(0) \left[\frac{M_s(T)}{M_s(0)}\right]^n$ (Equation 5)

By replacing successively $H_u(T)$ (using Equation 4) and $K_u(T)$ (using Equation 5) in the in Equation 3, we obtain:

$$M_{eff}(T) = M_s(T) - \frac{2K_u(0)}{\mu_0 M_s(T)}\left[\frac{M_s(T)}{M_s(0)}\right]^n$$

$$M_{eff}(T) = M_s(T) - \frac{2}{\mu_0}K_u(0) \cdot M_s(T)^{n-1} \cdot M_s(0)^{-n}$$

And finally:

$$M_{eff}(T) = M_s(T)\left[1 - \frac{2}{\mu_0}K_u(0) \cdot M_s(T)^{n-2} \cdot M_s(0)^{-n}\right] = M_s(T) \cdot A_{n,T}$$

The temperature dependence is obtained by calculating :

$$\frac{\partial M_{eff}(T)}{\partial T} = M_s(T) \cdot \frac{\partial A_{n,T}}{\partial T} + A_{n,T} \cdot \frac{\partial M_s(T)}{\partial T}$$

All terms involved in defining $A_{n,T}$ being temperature independent but the saturation magnetization, a vanishing temperature dependence of the effective magnetization can only be obtained by having $n = 2$ :

$$\frac{\partial A_{n,T}}{\partial T} = \frac{\partial}{\partial T}\left(1 - \frac{2}{\mu_0} K_u(0) \cdot M_s(T)^0 \cdot M_s(0)^{-2}\right) = 0$$

Furthermore, fixing $n = 2$ allows to write :

$A_{2,T} = 1 - \frac{H_u(0)}{M_s(0)} \approx 0$ when we achieve vanishing effective magnetization.

As a conclusion, obtaining a temperature independent effective magnetization requires two strong conditions : a Callen and Callen exponent that is equal to 2 and a magnetic anisotropy that compensates almost exactly the dipolar field.